\documentclass[conference]{IEEEtran}
\IEEEoverridecommandlockouts
\usepackage{cite}
\usepackage{amsmath,amssymb,amsfonts}
\usepackage{algorithmic}
\usepackage{graphicx}
\usepackage{textcomp}
\usepackage{xcolor}
\usepackage{amsmath}
\usepackage{amsfonts}
\usepackage{url}
\usepackage{array}  
\def\BibTeX{{\rm B\kern-.05em{\sc i\kern-.025em b}\kern-.08em
    T\kern-.1667em\lower.7ex\hbox{E}\kern-.125emX}}
\begin{document}

\title{Towards Bridging Formal Methods and Human Interpretability: Evaluating Human Comprehension of Labeled Transition System Designs}


\author{\IEEEauthorblockN{Abhijit Paul}
\IEEEauthorblockA{\textit{IIT, University of Dhaka, Bangladesh} \\
bsse1201@iit.du.ac.bd}
\and
\IEEEauthorblockN{Proma Chowdhury}
\IEEEauthorblockA{\textit{IIT, University of Dhaka, Bangladesh} \\
bsse1132@iit.du.ac.bd}
\and
\IEEEauthorblockN{Kazi Sakib}
\IEEEauthorblockA{\textit{IIT, University of Dhaka, Bangladesh} \\
sakib@iit.du.ac.bd}
}

\maketitle

\begin{abstract}
Labeled Transition Systems (LTS) are integral to model checking and design repair tools. System engineers frequently examine LTS designs during model checking or design repair to debug, identify inconsistencies, and validate system behavior. Despite LTS’s significance, no prior research has examined human comprehension of these designs. To address this, we draw on traditional software engineering and graph theory, identifying 7 key metrics: cyclomatic complexity, state space size, average branching factor, maximum depth, Albin complexity, modularity, and redundancy. We created a dataset of 148 LTS designs, sampling 48 for 324 paired comparisons, and ranked them using the Bradley-Terry model. Through Kendall's Tau correlation analysis, we found that Albin complexity (\(\tau = 0.444\)), state space size (\(\tau = 0.420\)), cyclomatic complexity (\(\tau = 0.366\)), and redundancy (\(\tau = 0.315\)) most accurately reflect human comprehension of LTS designs.
To showcase the metrics' utility, we applied the Albin complexity metric within the Fortis design repair tool, ranking system redesigns. This ranking reduced annotators' comprehension time by 39\%, suggesting that metrics emphasizing human factors can enhance formal design interpretability. 
\end{abstract}


\section{Introduction}

The rise of artificial intelligence (AI), particularly large language models (LLMs), has transformed software development. This has placed a new emphasis on precise software specifications. LLMs rely on accurate specifications to generate the desired code. However, the creation and comprehension of accurate specifications remains a significant challenge. This challenge is particularly acute in the context of formal specifications, where the complexity of visual representations often creates barriers to effective validation and debugging.

Visualization tools have emerged as a potential solution to this challenge. Tools such as the Labeled Transition System Analyzer (LTSA) \cite{magee2006concurrency} and Alloy \cite{jackson2012software} incorporate built-in visualization capabilities to aid in specification analysis. Automated design repair tools like Fortis and Oasis detect and repair specification issues, optimizing for cost, minimal change, and safety. However, these tools often neglect human comprehension, producing designs that, while technically correct, are overly complex and impractical. This highlights a crucial gap - the need to incorporate human comprehension metrics into the 
algorithms that drive these tools.



This study addresses these gaps by investigating design metrics that can effectively approximate human comprehension of software designs. The study is specifically focused on Labeled Transition System (LTS) designs because of their widespread use for prominent model-checking tools such as SPIN
, NuSMV
, LTSA and design repair tool like Fortis.
We considered design complexity and structural metrics 
to evaluate LTS designs in this work. Metrics like Albin complexity \cite{Albin1980complexity}, cyclomatic complexity \cite{edmonds1999syntactic}, state space size \cite{holzmann1997spin} were used to design complexity. For structural metrics, we found redundancy \cite{costa2021further}, modularity \cite{newman2006modularity}, average branching factor, max depth appropriate for this work. 
\newline To evaluate the effectiveness of these 7 metrics in approximating human comprehension, 
we construct a dataset of LTS designs derived from Kramer's book \cite{magee_kramer_2006} and Fortis-generated redesigns. The dataset initially had 148 LTS designs.  Since ranking the designs by pairwise comparisons has shown improved reliability over ordinal classification \cite{jang2022decreasing}, we use pairwise comparison on 48 randomly sampled designs since \(\binom{148}{2}\) is a large number. 3 annotators annotated 324 pairs of designs, sampled from \(\binom{48}{2}\) or 1128 pairs. The agreement score of pairwise comparison between 3 annotators was 75.748\%. 
The resulting paired comparison data is then used to rank the designs using the Bradley-Terry model \cite{hunter2004mm}. \newline
We used Kendall's Tau \cite{kendall1938rank} to measure the correlation between metric-based rankings and human-derived rankings. Results indicate that Albin complexity, state space size, redundancy, and cyclomatic complexity align most effectively with human comprehension of LTS designs. We further validated our approach through a case study. For our case study, we implemented the best-performing metric, Albin complexity, on top of Fortis and generated 16 repaired designs for a simple voting system described in \cite{zhang2023fortis}. To evaluate its impact, we divided the redesigns into two groups: one where the 16 designs were presented in their original order, and another where the designs were ranked based on Albin complexity to align with human comprehension. Two annotators then explored these designs to understand the repairs. We observe that the annotator who explored the ranked designs, moving from easier to more complex designs, completed their exploration 39\% faster.

\section{Related Works} \label{section: litreview}
In this section, we review existing literature across three key dimensions: technical design evaluation, human-centric design assessment, and graph complexity analysis, as LTS is fundamentally a graph-based representation. We also review design repair literature to emphasize the research gap on human-centric design assessment. 

\subsection{Evaluation of Software Design from Technical Perspective} \label{subsection: technical_design_metric}
Software design evaluation has traditionally focused on technical metrics, with limited consideration for design comprehension. The primary evaluation criteria typically encompass maintainability, performance, and reusability. Butler et al. propose management and maintenance coefficients based on McCabe's cyclomatic complexity metric \cite{butler2021metric}. While their work primarily addresses architectural risk analysis and statistical outlier detection, they suggest that McCabe's cyclomatic complexity (v) serves as an intuitive measure of design comprehension—as v increases, both understandability and maintainability tend to decrease. Pressman emphasizes modularity as a key metric \cite{pressman2005software}, while Sommerville focuses on cohesion and coupling \cite{sommerville2011software}. Bass et al. highlight the importance of scalability in design assessment \cite{bass2012software}, and the IEEE standards emphasize reusability as a critical measure \cite{ieee2000ieee}. 
Almansour's dEv framework \cite{almansour2014promoting} takes a more integrated approach, incorporating design evaluation into the SDLC through questionnaires, focus group discussions, think-aloud sessions, and heuristic evaluation. This framework represents a step toward more comprehensive design assessment, though it still primarily focuses on technical aspects rather than comprehension.


\subsection{Evaluation of Software Design from Human Perspective} \label{subsection: hci_design_metric}
Research on human comprehension of software design has primarily focused on visualization and layout aesthetics. Yusuf et al. \cite{yusuf2007assessing} assess UML diagram comprehension through eye gaze tracking, though this approach proves impractical for routine design evaluation as it requires direct user interaction.

Bergström et al. \cite{BERGSTROM2022111413} developed a more systematic approach by identifying properties affecting UML layout aesthetics. Their research revealed significant correlations between comprehension and several key factors, including the longest line, number of lines, rectangle orthogonality, number of rectangles, average line length, and line length variation. While LTSA employs automatic layout algorithms to enhance readability through minimal line crossing and balanced node distribution, our study focuses on the inherent comprehension complexity of designs rather than their visual presentation.
The literature consistently emphasizes user engagement in design evaluation, as evidenced by the work of Hart et al. \cite{hart2015investigating}, Afiaz et al. \cite{afiaz2023evaluation}, Sutcliffe \cite{sutcliffe2010user}, and Kushniruk \cite{kushniruk2016participatory}.  
Additionally, while there are works on using LLM to generate UML from requirements \cite{de2024evaluating}, there are no previous works on using LLM to evaluate software designs to the best of our knowledge.

\subsection{Evaluating Complexity of Graphs} \label{subsection: graph_litrev}
Most works on LTS complexity focuses on state explosion problems as alphabet size grows \cite{cece2013three}, aiming to optimize and approximate solutions \cite{yu2015approximate, devillers2023complexity}. Additionally, the evaluation of labeled transition system (LTS) designs relies on several key design metrics that assess their performance i.e. State Transfer Mechanisms, Self-Healing Properties \cite{bjrn_richerzhagen_2019} and behavioral characteristics i.e. Isomorphism Equivalence , Trace Equivalence, Bisimulation Equivalence \cite{roberto_gorrieri_2017}. These approaches, however, do not consider traditional design metrics discussed in the technical design metric section, let alone human comprehension metrics.

Edmond and Bruce analyzes complexity of a wide range of formal modeling languages and reports that:
\begin{itemize}
    \item Expressions with no repetitions are simple
    \item Small size should limit the possible complexity
    \item The Cyclomatic Number as a Measure of Analytic Complexity \cite{edmonds1999syntactic}
\end{itemize}

Given the absence of LTS-specific complexity metrics from a software design perspective, we review graph theory, as LTS are essentially directed graphs. Albin et al shows that the complexity of each nodes depends upon the order of the graph, the degree of the node and the longest path parameter of the graph. The combined complexity of nodes constitutes the complexity of the graph \cite{Albin1980complexity}. 
For estimating traditional software design metrics, particularly modularity and redundancy, we employ novel approaches. Modularity can help determine how well the nodes can be divided into distinct regions that exhibit high internal connectivity and low external connectivity. 

Newman et al proposes an algorithm to detect community structure in LTS \cite{newman2004detecting}. We leverage it to approximate modularity in LTS. 
Additionally, we can identify redundancy by checking for states that have identical outgoing transition structures (i.e., states that have the same set of reachable states). 

\subsection{Design Repair Tools} \label{subsection: design_repair}
Existing design repair approaches typically neglect human comprehension factors. Buccafurri et al. formalize design repair through abductive model revision, focusing on elementary corrections of Boolean expressions to reduce the search space, further optimized by counterexamples \cite{buccafurri1999enhancing}. While their approach effectively generates repairs, it does not account for human interpretability. Similarly, Menezes et al. propose a method for realistic repair generation that considers domain constraints and actions driving state transitions, though it also lacks focus on human factors in design understanding \cite{de2010system}.
Chatzieleftheriou et al. address the state explosion problem with an abstraction refinement technique for model repair \cite{chatzieleftheriou2015abstract}. Additionally, Fortis generates redesigns based on cost factors and minimal change to meet evolving requirements and environments, though minimal change here also does not correspond to ease of human comprehension \cite{zhang2023fortis}. These approaches, while technically sound, demonstrate the persistent gap between automated repair capabilities and human-centered design considerations.

\section{Methodology} 
\begin{figure}
    \centering
    \includegraphics[width=\linewidth]{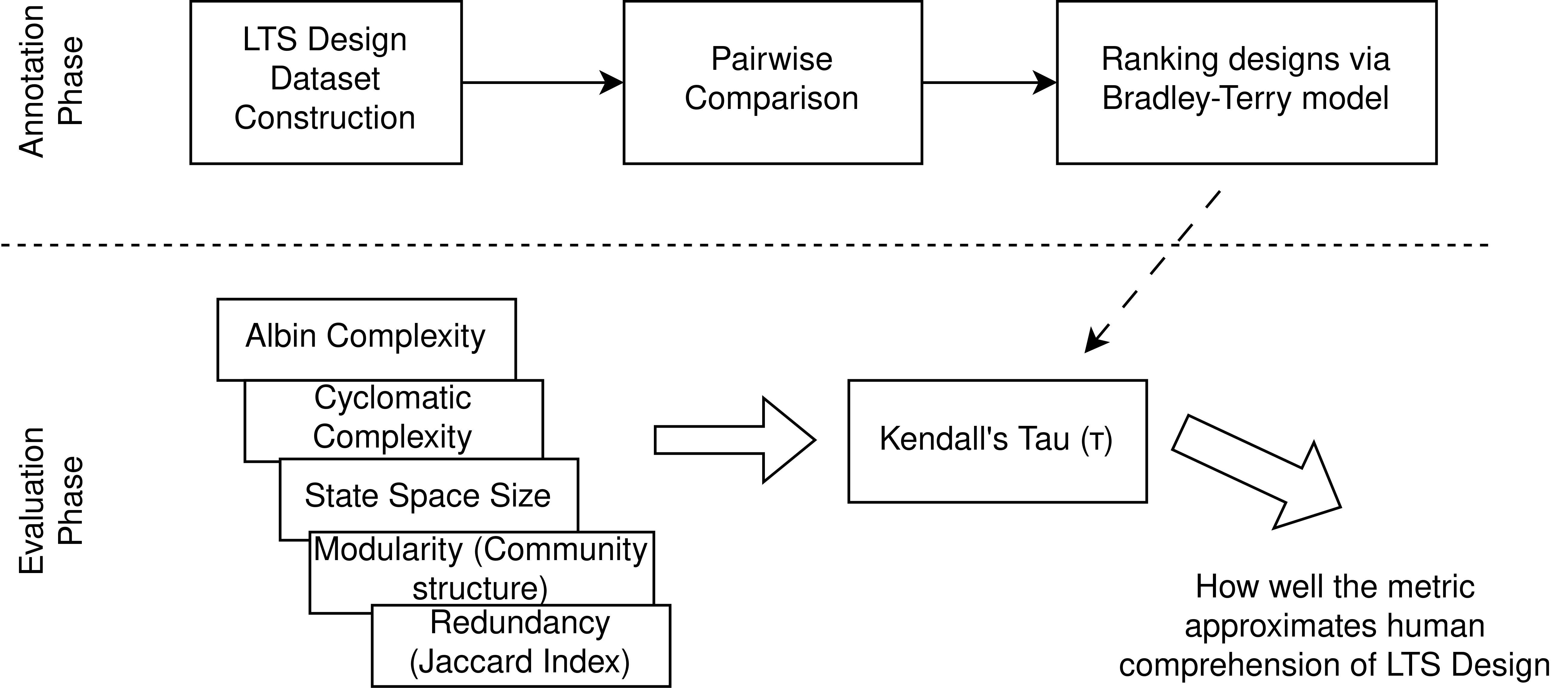}
    \caption{Process to evaluate the metrics}
    \label{fig:methodology}
\end{figure}

To evaluate how well the identified design metrics can approximate human comprehension in LTS design, we need a human annotated dataset of LTS designs where annotators rank the designs based on perceived complexity. Hence, our methodology is comprised of three sections: dataset construction, annotation and finally experimentation. More specifically, we first construct dataset of LTS designs. We then annotate it using pairwise comparisons. Based on these comparisons, we rank the designs using Bradley-Terry model \cite{hunter2004mm}. We then estimate the metrics identified in literature review (Section \ref{section: litreview}) and evaluate how they approximate human annotation using Kendall's Tau \cite{kendall1938rank}.  The methodology is summarized in Figure~\ref{fig:methodology}.


\subsection{Dataset Construction} \label{subsection:dataset_construction}
To the best of our knowledge, there is no publicly available LTS dataset. To bridge this gap, we collected LTS designs from two sources namely Kramer's book\cite{magee_kramer_2006} and the case studies of the Fortis tool\cite{zhang2023fortis}. 

We collected 97 specifications from Kramer's book \cite{magee_kramer_2006}. The specifications are defined in the Finite State Process algebra. We utilized LTSA to extract the LTS (.aut) representation of each specification, subsequently expressing them in matrix notation to leverage their graph-based properties. 
However, for systems consisting of multiple designs, 
their parallel composition were too large and complex. Since these compositions primarily serve model checking purposes rather than human evaluation, we collected individual designs instead of their composition for human annotation. We supplemented this dataset with 4 system specifications from Fortis's case study \cite{zhang2023fortis}. Through the application of Fortis, we generated additional redesigns of these systems, yielding 51 
designs.

\subsection{Annotation Process}\label{subsection:annotation}
The annotation process began with a manual inspection of 148 specifications. We observed that some were too complex to understand without any prior knowledge of the system. This initial review led us to eliminate 76 specifications. We then generated LTS designs in png format for the remaining 72 specifications using LTSA. 


However, annotating the design complexity of a diagram is a subjective task and hence, prone to errors. So we instead take inspiration from psychometrics and cognitive psychology. According to Jang et al \cite{jang2022decreasing}, ranking by pairwise comparisons has shown improved reliability over ordinal classification. This approach has also been used to assess preferences or perceived complexity between design alternatives \cite{yannou2009fixing}. So we find the paired comparison method \cite{david1963method} to be suitable for our case. 

Our annotation tool 
enables participants to compare pair of designs directly side by side, with zoom functionality for detailed examination. Participants are asked to assess which design in each pair is preferred in terms of complexity. The tool records both the time spent examining each design and the total decision-making duration. Given the computational complexity of evaluating all possible pairs \(\binom{48}{2}\) or 1128, we constrained our annotation to 324 samples. 3 annotators annotated the same 324 pairs of designs. The agreement score was 75.748\%.

For ranking derivation from pairwise comparisons, we employed the Bradley-Terry model \cite{hunter2004mm, caron2012efficient}. While we considered alternative approaches such as normalization and active ranking \cite{jamieson2011active}, initial manual inspection suggested superior performance from the Bradley-Terry model. However, we acknowledge the need for more rigorous validation of this selection in future work.


\subsection{Experimentation} \label{section:methodology}
From our literature review in Section \ref{section: litreview},  we estimate the following metrics based on\cite{Albin1980complexity, edmonds1999syntactic, mateos2022graph}. We divide the design metrics in two parts - complexity and structure based.
\subsubsection{Design Complexity Metrics} These metrics are inspired from traditional software engineering. Our intuition was that complexity measures can approximate design comprehension difficulty.

\textbf{Cyclomatic Complexity}
McCabe's cyclomatic complexity \textbf{\textit{V}} for a design measures the number of linearly independent paths through its state transition graph, indicating the design's structural complexity. It provides an intuitive measure of design comprehension: as \textbf{\textit{V}} increases, understandability and maintainability tend to decline \cite{butler2021metric, edmonds1999syntactic}. So we implement McCabe's cyclomatic complexity to measure design comprehension.
\[
V = E – N + 2P
\]
where E corresponds to edges, N to nodes, and P to connected components.

\textbf{Albin Complexity:} Albin et al\cite{Albin1980complexity} shows that the complexity of each node automaton depends on the order of the graph, the degree of the node and the longest path parameter of the graph. The combined complexity of nodes constitutes the complexity of the graph. We indicate this notion of complexity using "Albin Complexity" (the name of the author) to differentiate it from other complexity measures we use.
\[
\text{Albin Complexity} = n + \sum_{i=1}^n \text{deg}(v_i) + L
\]
where:
\begin{itemize}
    \item \(n\) is the order of the graph (the total number of nodes),
    \item \(deg(vi)\) is the degree of each node \(v_i\),
    \item \(L\) is the longest path parameter (in terms of edge count) in the graph.
\end{itemize}

\textbf{State Space Size:} The number of states can serve as a straightforward measure of complexity \cite{holzmann1997spin}. Large state spaces increase the cognitive load for users attempting to understand the system's behavior. Simplified state representations (fewer states and transitions) are often considered more user-friendly \cite{valmari1996state, edmonds1999syntactic}. We also consider max depth and average branching factor because deep sequences or large branching factors are typically more complex due to the compounded number of paths and possible interactions.

\subsubsection{Structural Metrics} We employ two key structural metrics: redundancy and modularity to evaluate LTS design. We estimate them for LTS by drawing from graph theory and set similarity. We call these design metrics structural metrics because they analyze the fundamental organization and composition of the LTS design 

\textbf{Modularity:} Designs with modular structures may be easier to understand and analyze compared to flat or monolithic designs \cite{parnas1972criteria}. It is because system engineers can focus on region, understand it and then, go to the next region. It becomes similar to a divide and conquer strategy. Modularity is a measure often used in network analysis to identify the strength of division of a network into modules (or communities). We can represent the LTS as a adjacency matrix. Modularity can help us determine how well the nodes can be divided into distinct regions that exhibit high internal connectivity and low external connectivity. We use the following algorithm to detect community structure in our LTS \cite{newman2004detecting, newman2006modularity}. It includes the following steps:
\begin{enumerate}
    \item Express the LTS into an adjacency matrix $A$, where:
    \begin{itemize}
        \item \(A[i][j] = 1\)  if there is a connection between nodes \(i\) and \(j\).
        \item \(A[i][j] =  0\) if there is no connection between nodes \(i\) and \(j\).
    \end{itemize}

    \item Now we calculate the degree f each node. The degree $k_i$ of a node $i$ is the sum of the entries in row $i$ of the adjacency matrix:
        $$k_i = \sum_j A[i][j]$$  
    \item The total number of edges $E$ in the graph can be computed as:

        $$E = \frac{1}{2} \sum_i k_i$$
    
    This counts each edge twice (once for each node).

    \item To detect community structure in the graph, apply the Girvan-Newman algorithm \cite{newman2004detecting}, which iteratively removes edges with high betweenness to reveal communities. As the edges are removed, the graph breaks into separate components that are identified as communities.
    
    \item Once the communities are identified, assign community labels to each node.
    \item The modularity $Q$ is defined as \cite{newman2006modularity}:

    $$Q = \frac{1}{2E} \sum_{i,j} \left(A_{ij} - \frac{k_i k_j}{2E}\right) \delta(c_i, c_j)$$
    
    Where:
    \begin{itemize}
        \item $A_{ij}$ is the entry of the adjacency matrix indicating the presence of an edge between nodes $i$ and $j$.
        \item $k_i$ and $k_j$ are the degrees of nodes $i$ and $j$.
        \item $c_i$ and $c_j$ are the communities (or modules) to which nodes $i$ and $j$ belong.
        \item $\delta(c_i, c_j)$ is 1 if $i$ and $j$ are in the same community and 0 otherwise.
    \end{itemize}

    \item After identifying the communities and assigning the labels, substitute these community labels into the modularity formula to compute \(Q\).
\end{enumerate}









\textbf{Redundancy:} Redundancy adds to complexity, as they can obscure the unique behaviors of the model. Techniques like symmetry reduction are used to minimize this redundancy and simplify analysis \cite{emerson1996symmetry}. As discussed in section \ref{subsection: graph_litrev}, we here use Jaccard index to find similarity between two states \cite{costa2021further}. For two nodes \(u\) and \(v\) from LTS, the Jaccard Index \(J(u,v)\) is defined as:
\[
J(u, v) = \frac{|N(u) \cap N(v)|}{|N(u) \cup N(v)|}
\]
Where \(N(u)\) represents the set of neighbors of node \(u\), and  \(N(v)\) represents the set of neighbors of node \(v\).


\section{Evaluation}
We evaluate the effectiveness of the 7 metrics in approximating human comprehension. Additionally, we explore whether consideration of human comprehension can improve user experience. For that, we focus on two specific research questions.

\textbf{RQ1 (Effectiveness):} Can traditional software design metrics approximate human comprehension of software design?

\textbf{RQ2 (Use Case):} Does accounting for human comprehension reduce the time required to understand designs in formal specification tools?

To address RQ1, we adopt a quantitative approach by calculating the Kendall's Tau \cite{kendall1938rank} correlation between metric-based rankings and human-derived rankings. For RQ2, we take a qualitative approach by implementing our metrics in a design repair tool called Fortis, and evaluating whether users can navigate through the generated redesigns more quickly.

\subsection{Effectiveness of metrics} \label{sec: result}
We assess the effectiveness of the 7 metrics to approximate human comprehension using Kendall's Tau (\(\tau\)). Kendall's Tau is a correlation measure suitable for evaluating the relative orderings of our design rankings. This approach aligns with our goal of capturing the relative complexity of designs across the entire set of 48 items rather than pinpointing exact positions. Table~\ref{table:result} summarizes our findings.  

A higher Kendall's Tau \(\tau\) value indicates a stronger alignment with human judgment, whereas a value closer to zero suggests weak or no correlation. For example, metrics such as State Size, Redundancy and Albin Complexity exhibit a high correlation with human ratings, demonstrating their effectiveness in reflecting perceived design complexity. Conversely, Modularity shows a lower correlation (-0.236), indicating that Newman's algorithm \cite{newman2004detecting} for community structure in a network may not be suitable to measure modularity in LTS.

\begin{table}[ht]
    \centering
    \caption{Statistical Correlation Between Design Metrics and Human Comprehension Rankings}
    \label{table:result}
    \begin{tabular}{|l|r|r|}
    \hline 
Design Metric             & Kendall's Tau (\(\tau\)) & P-value  \\ \hline
Cyclomatic Complexity (V) \cite{edmonds1999syntactic} & 0.3656565657  & 0.0003983119023 \\ \hline
State Space Size \cite{holzmann1997spin}                & 0.4202020202  & 4.71E-05        \\ \hline
Average Branching Factor  & 0.09898989899 & 0.3377263584    \\ \hline
Max Depth                 & 0.4202020202  & 4.71E-05        \\ \hline
Albin Complexity \cite{Albin1980complexity}          & 0.4444444444  & 1.68E-05        \\ \hline
Modularity (Q) \cite{newman2004detecting}            & -0.2363636364 & 0.02207570243   \\ \hline
Redundancy (J) \cite{costa2021further}       & 0.3151515152  & 0.00227258672   \\ \hline
    \end{tabular}
\end{table}

The corresponding p-values provide more insights into the statistical significance of these correlations. 
Metrics like Cyclomatic Complexity, State Size, Redundancy, and Albin Complexity have very small p-values (below 0.01), suggesting that the observed correlations are statistically significant and not due to random chance. On the other hand, the Average Branching Factor and Modularity yield higher p-values, indicating weaker or non-significant relationships with human rankings. These results underscore the utility of specific design metrics (e.g., State Size and Albin Complexity) in predicting human assessments of design complexity, while also revealing the limitations of others like Modularity in this regard.

\subsection{Use Case Study} \label{section: case-study}
To evaluate the practical effectiveness of our findings, we conducted a case study on the tool Fortis \cite{zhang2023fortis}. We implemented\footnote{\url{https://github.com/abj-paul/Robustify-Design}} the Albin Complexity metric—the highest-performing metric in our study—on top of Fortis to rank generated redesigns. Fortis is a design repair tool that, given a system’s behavioral models, environmental models, and a set of user-specified deviations, produces one or more redesigns capable of satisfying desired properties despite environmental changes. By ranking the redesigns based on their Albin Complexity, we aim to determine whether this approach facilitates human comprehension of the generated designs.

\begin{figure}
    \centering
    \includegraphics[width=0.5\linewidth]{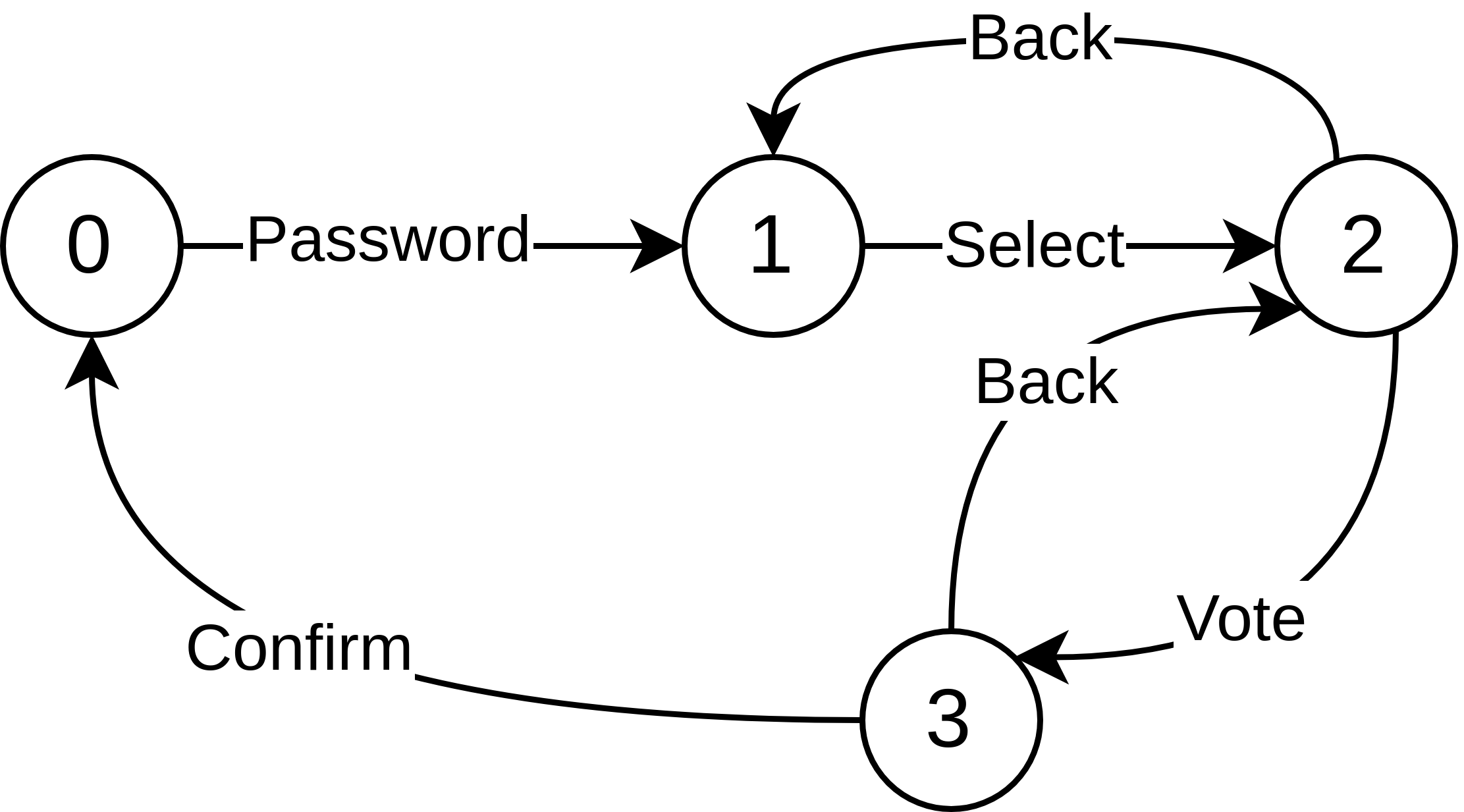}
    \caption{Voting System \cite{zhang2023fortis}}
    \label{fig:votingsystem}
\end{figure}

The experiment was conducted with two human annotators, assessing the time needed to understand the ranked and unranked redesigns. Using Fortis, we generated redesigns for a simple voting system (figure-\ref{fig:votingsystem}) as defined in \cite{zhang2023fortis}. We chose this system because, despite the simplicity of the voting mechanism, the 16 generated redesigns are considerably more complex, with an average state space size of 40.36 (standard deviation: 11.73). Furthermore, the voting system’s straightforward design made it easy to work on for our annotators, who are software engineers with two years of experience. These designs were divided into two groups: the first group presented the redesigns in their original, unranked order, while the second group ordered them from least to most complex based on their Albin Complexity scores. We hypothesized that the ranked ordering would reduce comprehension time by presenting simpler designs first, allowing annotators to incrementally build understanding before examining more complex designs.

The results supported this hypothesis: annotators required 66 minutes to navigate the unranked redesigns but only 40 minutes to comprehend the ranked set, a 39\% reduction in time. This initial case study suggests that ranking designs based on human comprehension metrics may enhance the usability and interpretability of automated design tools like Fortis. While a larger-scale study is necessary to further validate these results, these preliminary insights demonstrate potential benefits. Future work will investigate integrating comprehension-based metrics directly into optimization functions within design repair tools, potentially enabling the generation of more interpretable and user-friendly redesigns.

\section{Conclusion and Future Work}
In this work, we explore the human comprehension of LTS designs. We explore literature and identify 7 design metrics. We propose novel approach to estimate redundancy and modularity metrics for LTS design. We annotate a dataset of 48 designs using pairwise comparison. We find moderate correlation between metric-based rankings and human-derived ranking, suggesting that traditional software metrics can indeed approximate human comprehension. We show a case study on a simple voting system, highlighting that consideration of human comprehension can decrease 39\% time in exploring design alternatives. These findings show preliminary evidence that we should put more focus in human comprehension of designs generated by tools like LTSA, Fortis, Alloy. Designs that are easier to interpret are easier to debug, identify inconsistencies in, and understand. Our work can contribute to ongoing works on better visualization for formal specification and developing design repair tools optimized for human interpretability.

LTS are integral to model checking and design repair tools. System engineers frequently examine LTS designs during model checking or design repair to debug, identify inconsistencies, and validate system behavior. So it is necessary to consider human factors associated with LTS designs to make it easier for system designers to spot inconsistencies and navigate through the design alternatives. So in the future, we plan to expand this work by incorporating a larger dataset and exploring additional metrics and algorithms.
Additionally, we will explore the integration of human comprehension metrics into design repair tools, aiming to generate redesigns that are optimized for interpretability. 


\section{Threats To Validity}
Evaluating design complexity on a limited dataset of 48 LTS designs poses an external validity threat, as findings may vary with different datasets. To address this, we carefully curated designs from both academic literature and tool-generated modifications to capture diverse design patterns. However, the dataset may still not fully reflect the range of complexity found in real-world applications. Confounding factors in annotator experience and unfamiliarity with specific design patterns present an internal validity concern, since perceived difficulty can be due to ignorance. To mitigate individual bias, we utilized a pairwise comparison method with 3 annotators, although random sampling of pairs might miss certain complexity relationships. The source code and annotated dataset is provided to support evaluation reproducibility and transparency for future studies.

\bibliographystyle{IEEEtran}
\bibliography{main}

\end{document}